\documentclass[twocolumn,showpacs,showkeys,preprintnumbers,amsmath,amssymb]{revtex4-1}

\usepackage{graphicx}
\usepackage{dcolumn}
\usepackage{bm}

\begin{document}


\title[]{Threshold of auroral intensification reduced by electron precipitation effect}

\author{Yasutaka Hiraki}
 \email{yhiraki28@gmail.com}
\affiliation{
University of Electro-Communications, 1-5-1 Chofugaoka, Chofu, Tokyo, 182-8585, Japan.\\
}


\date{\today}

\begin{abstract}
It has been known that discrete aurora suddenly intensifies and deforms from an arc-like to a variety of wavy/vortex structures, especially during a substorm period. The instability of Alfv$\acute{\rm e}$n waves reflected from the ionosphere has been analyzed in order to comprehend the ignition process of auroral intensification. It was presented that the prime key is an enhancement of plasma convection, and the convection electric field has a threshold. This study examined effects of auroral electron precipitation, causing the ionization of neutral atmosphere, on the linear instability of Alfv$\acute{\rm e}$n waves. It was found that the threshold of convection electric fields is significantly reduced by increasing the ionization rate, the realistic range of which could be estimated from observed electron energy spectra. 
\end{abstract}

\maketitle

\section{Introduction}\label{sec: 1}
Auroral intensification has been vigorously studied with narrow field of view or high resolution optical measurements [e.g., Donovan et al., 2007; Mende et al., 2009]. It is known that the time scale of intensification is 1--2 min, and fading of auroral arcs appears just before the onset. Furthermore, vortex street structures called as "beads" emerged on the auroral arc, probably due to a development of strong flow shears. 

It has been considered that, on the basis of the above facts, shear Alfv$\acute{\rm e}$n waves propagating along the magnetic field line are involved in auroral intensification. Three-dimensional simulations that treated amplification of shear Alfv$\acute{\rm e}$n waves, driven by a coupling between the ionospheric and magnetospheric plasmas with field-aligned currents, were performed [Jia and Streltsov, 2014; Hiraki, 2015a, b]. Hiraki [2015b] presented that a threshold exists for the instability of Alfv$\acute{\rm e}$n waves due to an enhancement of plasma convection; the threshold electric field was $\approx 25$ mV/m in their situation. These studies did not explicitly treat the electron acceleration by field-aligned electric fields and the ionization of neutral atmosphere, which are the essence of discrete auroras. They assumed that the upward field-aligned current carried by waves is identical to the auroral luminosity. 

Measurements of electron density by EISCAT radar [Kirkwood et al., 1990; Olsson et al., 1996] and of particles by DMSP and FAST satellites [Yago et al., 2005; Mende et al., 2003] revealed that the field-aligned electric field originates from inverted-V type potential drops at a height of $\approx 1000$ km or inertial Alfv$\acute{\rm e}$n waves excited at $\approx 6000$ km. Energies of electrons accelerated by each electric field are 2--10 keV or $< 1$ keV, respectively. Although previous simulations treated inertial Alfv$\acute{\rm e}$n waves [Lysak and Song, 2008; Jia and Streltsov, 2014], evolution of waves with a realistic field-aligned electric field is not still resolved. 

This study examined the other topic, i.e., the effect of ionization by inverted-V type non-thermal electrons on the instability of shear Alfv$\acute{\rm e}$n waves. These electrons with an energy of 2--10 keV strongly increase the conductivities in a height of 100--120 km where horizontal currents that can couple with shear Alfv$\acute{\rm e}$n waves maximize [Fang et al., 2008]. 

The ionization rate by the non-thermal electrons was formulated as $q j_\parallel$, $q$: electron yielding coefficient, $j_\parallel$: field-aligned current by Alfv$\acute{\rm e}$n waves [Lysak and Song, 2002]. However, its effect on the growth of shear Alfv$\acute{\rm e}$n waves was not clarified. This study analyzed the linear instability of shear Alfv$\acute{\rm e}$n waves, including effects of ionization, in the magnetosphere-ionosphere (M-I) coupling system. We especially focused on the possibility that the threshold field value ($\approx 25$ mV/m) presented in our previous study is reduced to be a popular level in observations, e.g., 10--20 mV/m [Provan et al., 2004; Bristow and Jensen, 2007].

\section{Model Description}\label{sec: 2}
The target region of our analysis is a magnetic flux tube that involves an auroral arc with a width of $\approx 10$ km. Shear Alfv$\acute{\rm e}$n waves, propagating along the magnetic field right above the arc, flow the field-aligned current $j_\parallel$ into the ionosphere. It connects with horizontal currents caused by plasma convection. A situation is considered where the coupling of currents destabilizes shear Alfv$\acute{\rm e}$n waves, implying the initial brightening of aurora. 

The field line position ${\bm s}$ is defined as $s = 0$ at the ionosphere and $s = l$ at the magnetic equator. We consider a latitude of $70^\circ$ in the southern hemisphere, with the dipole $L$ value of $\approx 8.5$ and $l \approx 7\times10^4$ km. We take coordinates ${\bm x}(s)$ and ${\bm y}(s)$ orthogonal to each $s$: ${\bm x}$ points southward (poleward) and ${\bm y}$ points eastward at $s = 0$. We set a local flux tube, e.g., a square of ($l_\perp \times l_\perp$) at $s = 0$ and a rectangle of ($\approx 3300$ km $\times$ $\approx 1700$ km) at $s = l$ using $l_\perp \approx 70$ km and dipole metrics; see Hiraki and Watanabe [2011] for details. 

The dipole magnetic field is written as ${\bm B}_0$. The system has a convective electric field ${\bm E}_0$ that is applied poleward ($\parallel {\bm x}$) and is uniform in every $x$-$y$ planes. With perturbed electric and magnetic fields of ${\bm E}_1 = - B_0 {\bm \nabla}_\perp \phi$ and ${\bm B}_1 = {\bm \nabla}_\perp \psi \times {\bm B}_0$, linearized equations for shear Alfv$\acute{\rm e}$n waves are expressed as 
\begin{eqnarray}
 \displaystyle && \partial_t \omega + {\bm v}_0 \cdot {\bm \nabla}_\perp \omega = v_{\rm A}^2 \partial_s j_\parallel  \\
 && \partial_t \psi + {\bm v}_0 \cdot {\bm \nabla}_\perp \psi = - \frac{1}{B_0} \partial_s B_0 \phi. 
\end{eqnarray}
The domain of definition is $0 < s \le l$. Here, $\omega = \nabla_\perp^2 \phi$ stands for vorticity, $j_\parallel = - \nabla_\perp^2 \psi$ field-aligned current, ${\bm v}_0 = {\bm E}_0 \times {\bm B}_0 / B_0^2$ the convection speed, and $v_{\rm A}$ the Alfv$\acute{\rm e}$n velocity. The same as Hiraki [2015b], $v_{\rm A}$ is set to be constant as $\approx 1.5 \times 10^3$ km/s along the field line. The Alfv$\acute{\rm e}$n transit time is thus $\tau_{\rm A} = \int_{0}^{l} 1/v_{\rm A}(s) {\rm d}s \approx 47$ s. 

By integrating the continuity equations of ions and electrons, the linearized equations 
\begin{eqnarray}
 \displaystyle && \partial_t n_{\rm e} + {\bm v}_0 \cdot {\bm \nabla}_\perp n_{\rm e} = q j_\parallel - R n_{\rm e} \\
 && - n_0 \mu_{\rm P} \nabla_\perp^2 \phi + (\mu_{\rm P} {\bm E}_0 - {\bm v}_0) \cdot {\bm \nabla}_\perp n_{\rm e} = D \nabla_\perp^2 n_{\rm e} - j_\parallel
\end{eqnarray}
are obtained for plasma motion at the ionosphere $s = 0$. Here, $n_{\rm e}$ stands for electron density, $n_0$ its ambient component, $\mu_{\rm P}$ Pedersen mobility, $R$ the recombination coefficient, and $D$ the diffusion coefficient. The Hall mobility appearing in Eqs.\ (3) and (4) is normalized to be unity. These equations couple with Eqs.\ (1) and (2) of shear Alfv$\acute{\rm e}$n waves through field-aligned current $j_\parallel$. The electron yielding rate was set to be $q = 1$ in Hiraki [2015b] by assuming continuity of the thermal electron flux. In this study, $q\ge 1$ is given on the supposition that ionization of neutral particles occurs due to the non-thermal electron precipitation. 

In order to solve Eqs.\ (1)--(4), field variables $\phi$, $\psi$, and $n_{\rm e}$ are expanded such as $\phi({\bm x}, s) = \tilde{\phi}(s) {\rm exp}^{{\rm i} ({\bm k}_\perp \cdot {\bm x} - \Omega t)}$ with complex frequency $\Omega$ and perpendicular wave number ${\bm k}_\perp$. The boundary condition at the magnetic equator is set to be $\partial_s \phi = 0$, or $\psi = 0$ (zero current), and shapes of eigenfunctions $\tilde{\phi}$ and $\tilde{\psi}$ at $s = 0$--$l$ are calculated. We can redefine the frequency $\Omega - {\bm v}_0 \cdot {\bm k}_\perp$ in the frame of ${\bm v}_0$ to be $\Omega$ without loss of generality. From Eqs.\ (3) and (4), we get the dispersion relation as 
\begin{equation}
 \displaystyle \Bigl( 1 - q \frac{\sigma - {\rm i} D k_\perp^2}{\Omega + {\rm i} R} \Bigr) \tilde{\psi} + \alpha \tilde{\phi} = 0. 
\end{equation}
Here, two parameters $\sigma = (\mu_{\rm P} {\bm E}_0 - {\bm v}_0) \cdot {\bm k}_\perp$ and $\alpha = \mu_{\rm P} n_0$ are introduced. The former means the convective drift frequency, while the latter corresponds to the Pedersen/Alfv$\acute{\rm e}$n impedance ratio $\Sigma_{\rm P}/\Sigma_{\rm A}$. 

The ionospheric parameters used for the solution of Eq.\ (5) are set to be $\mu_{\rm P}$/$\mu_{\rm H} = 0.5$, $\alpha = 5$, $R = 2 \times 10^{-3}$ s$^{-1}$, and $D = 4 \times 10^5$ m$^2$/s. It is also noted that the ambient density and magnetic field are set to be $n_0\approx 3.8 \times 10^4$ cm$^{-3}$ and $B_0 \approx 5 \times 10^{-5}$ T.

\begin{figure}[t]
\includegraphics[width=1.0\columnwidth, bb=0 0 360 252, clip]{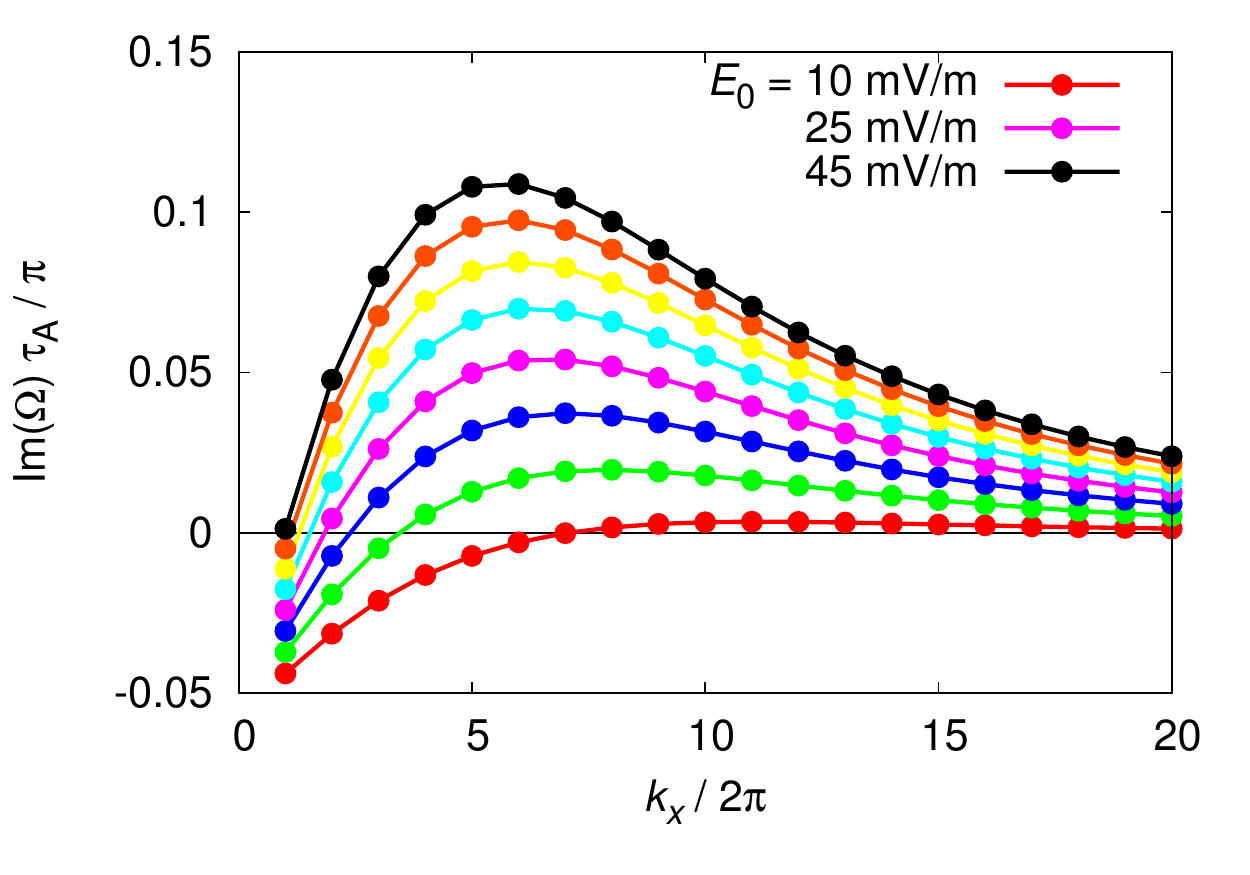}
\caption{Growth rate ${\rm Im}(\Omega)$ of finite $k_x$ and $k_y = 0$ modes, normalized by $\pi/\tau_{\rm A}$, in the nominal case of the electron yielding coefficient $q = 1$. The convection electric field is changed to be $E_0 = 10$ (red), 15, 20, 25 (purple), 30, 35, 40, and 45 (black) mV/m.}
\end{figure}

\begin{figure}[t]
\includegraphics[width=1.0\columnwidth, bb=0 0 360 252, clip]{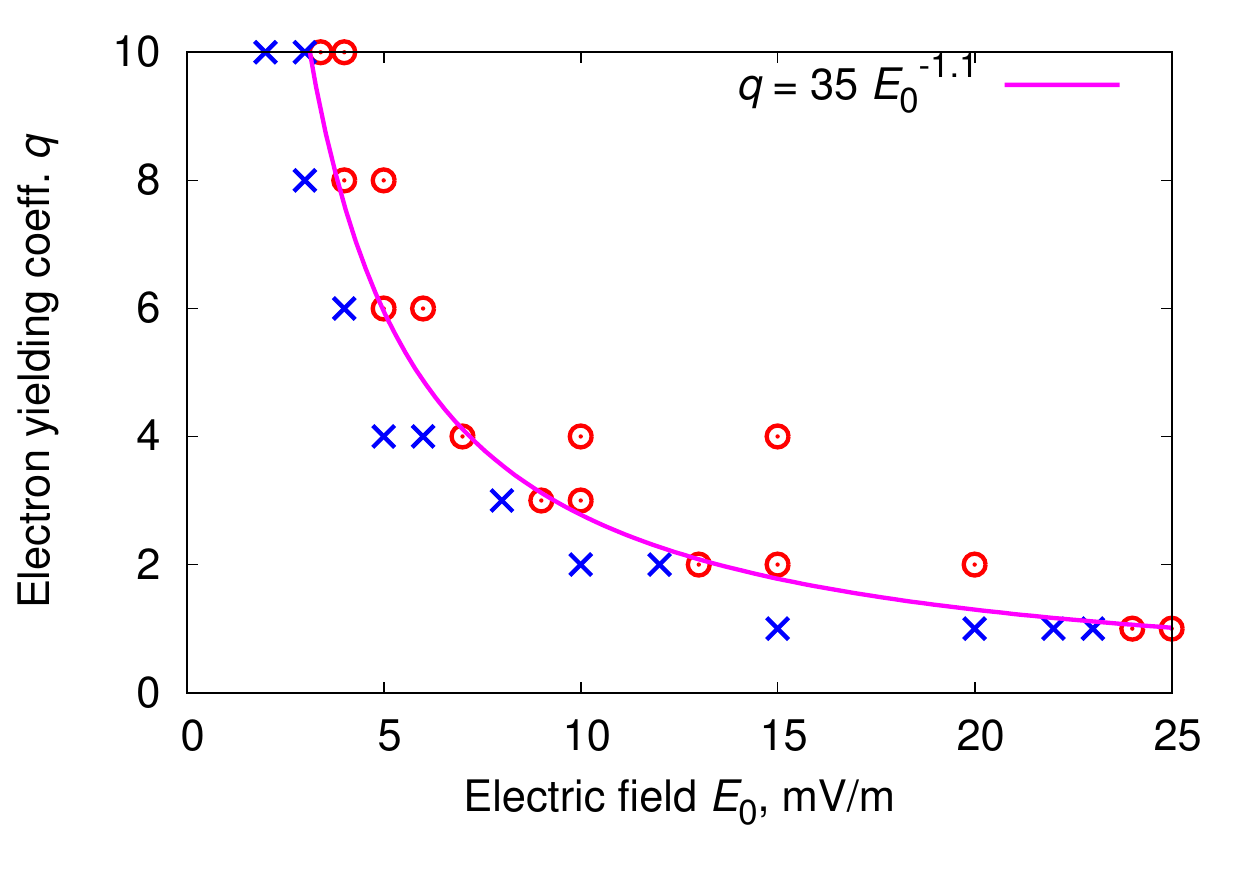}
\caption{A $E_0$-$q$ diagram of the growth rate $\gamma$ of $(k_x, k_y) = (2, 0)$ mode; open circles show the cases of $\gamma (2, 0)>0$ and crosses show the cases of $\gamma (2, 0) < 0$. A function $q = 35 E_0^{-1.1}$ is fitted to the open-cross boundary.}
\end{figure}

\begin{figure}[t]
\includegraphics[width=1.0\columnwidth, bb=0 0 410 770, clip]{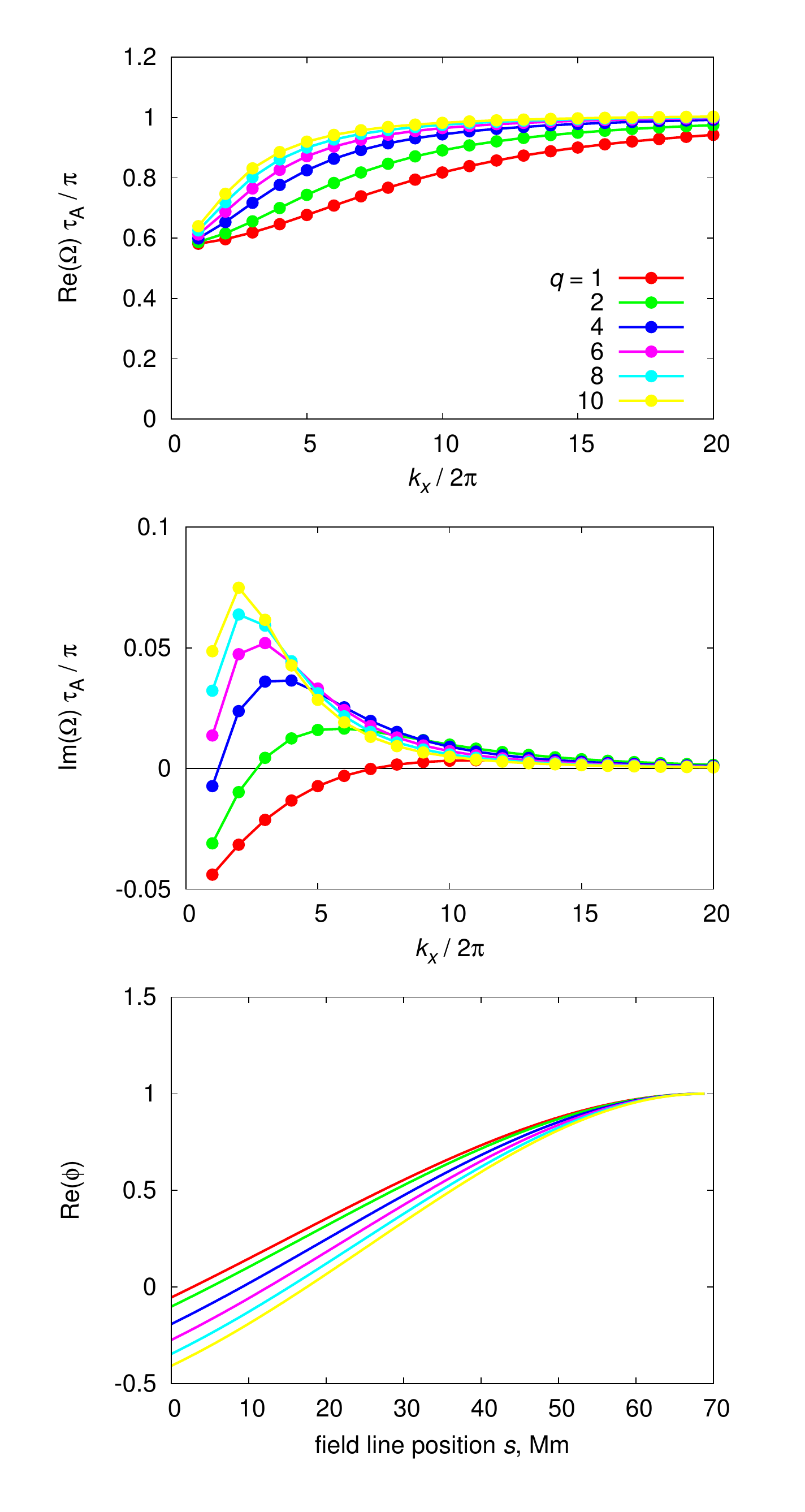}
\caption{Dependence of (a) real frequency ${\rm Re}(\Omega)$, (b) growth rate $\gamma$, and (c) eigenfunctions ${\rm Re}(\tilde{\phi})$ on the electron yielding rate $q$ for $(k_x, k_y) = (2, 0)$ mode along with $E_0 = 10$ mV/m.}
\end{figure}

\section{Results}\label{sec: 3}
The simplest case is that there are an auroral arc in the ionosphere and Alfv$\acute{\rm e}$n waves right above it. The $k_x = 1$ mode is distinguished in the arc if its spatial scale is approximated to the Gaussian distribution; hereafter, wave number is normalized by $2 \pi / l_\perp$. The arc without east-westward structures, i.e., $k_y = 0$, is considered for simplicity. Figure 1 shows the linear growth rates $\gamma \equiv {\rm Im}(\Omega) \tau_{\rm A} / \pi$ of ($k_x \ne 0$, $k_y = 0$) modes that were obtained by solving Eqs.\ (1), (2), and (5) under the conditions in Sec.\ \ref{sec: 2}. Here, the electron yielding rate being $q = 1$, the response of $\gamma$ to changes in the poleward electric field $E_0$ was found. We readily found the value of the critical field for $\gamma(1, 0) > 0$ to be $\approx 45$ mV/m. 

Hiraki [2015b] demonstrated that an arc quickly splits into two arcs due to a density perturbation produced by the poleward Pedersen current. For the higher mode $(2, 0)$, we also evaluated the critical field for $\gamma(2, 0) > 0$ to be $\approx 23$ mV/m (Fig.\ 1). It is, however, noted that a lower electric field, e.g.\ 10 mV/m, just before auroral intensification was estimated with SuperDARN radars [Provan et al., 2004]. 

We investigate how the critical field for this (2, 0) mode responds when $q$ ($\ge 1$) is changed. Figure 2 shows a diagram of $\gamma(2, 0)$ in ($E_0$, $q$) space: open circle stands for the case of $\gamma(2, 0) > 0$, while cross for the case of $\gamma(2, 0) < 0$. It is clearly found that the critical value of $E_0$ for $\gamma(2, 0) > 0$ can be reduced if $q$ is increased. Note that the term proportional to $q \sigma \sim q E_0$ appears in the left-hand side of Eq.\ (5) in case of $D = 0$. It means that a mode is unstable for any set of $q$ and $E_0$ that satisfies $q E_0 = {\rm const}$; the diffusion term ($D$) slightly changes the relation to a fitting function $q E_0^{1.1} = {\rm const}$ for the open circle-cross boundary in Fig.\ 2. Because the $q$ value has an upper limit in the real situation just before a faint arc starts to intensify, the critical $E_0$ necessarily has a lower limit. We will discuss in Sec.\ \ref{sec: 4} on the way to estimate the realistic range of $q$ values and the reason why we remove the possibility of higher modes $k_x \ge 3$. 

At the end of this section, we describe the characteristics of eigenmodes in the moderate case of electric field $E_0 = 10$ mV/m. Figure 3 shows the responses of real frequency ${\rm Re}(\Omega)$, growth rate ${\rm Im}(\Omega)$, and eigenfunction ${\rm Re}(\tilde{\phi})$ of the $(2, 0)$ mode to a change in $q$. The gradient of ${\rm Re}(\Omega)(k_x)$ steepens in the lower $k_x$ side, owing to $q \sigma \sim q k_x$ in Eq.\ (5). Related to this change as $q$ increases, a sharp peak in $\gamma$ appears and moves to the lower $k_x$ side. A high $\gamma$ in the case of high $q$ is supported by the fact that the amplitude of eigenfunction ${\rm Re}(\tilde{\phi})$ at $s = 0$ increases.

\section{Discussion}\label{sec: 4}
Let us first mention why the $(k_x, k_y) = (2, 0)$ mode is chosen when the critical field $E_{\rm cr}$ for auroral intensification is proposed at the part of Fig.\ 2. As pointed in Secs.\ \ref{sec: 1} and \ref{sec: 3}, the critical value $E_{\rm cr} \approx 45$ mV/m for the $(1, 0)$ mode in case of $q = 1$ is likely to be much higher than the typical value in observations [Bristow and Jensen, 2007]. The growth rate increases with the mode number, and the critical field for $(8, 0)$ is $E_{\rm cr} \approx 10$ mV/m (Fig.\ 1). However, we should start a speculation that $(2, 0)$, $(3, 0), \dots$ modes are produced through successive splits of an arc. It may require a very long time for the $(8, 0)$ mode to emerge to cause an instability involving Alfv$\acute{\rm e}$n waves. We suppose that an arc with a width of $l_{\rm arc}$ ($\approx 10$ km), splitting into two arcs, requires to move at least a distance of $l_{\rm arc}$ by a drift in the critical field $v_{\rm 0, cr} \propto E_{\rm cr}$. Producing $N$-arcs requires a distance of $(N-1) l_{\rm arc}$. The condition for this $(N-1)$-splitting is that the time scale $\Delta \tau$ is shorter than the time scale $\tau_{\rm A}$ representing field changes. This relation is expressed as 
\begin{equation}
 \displaystyle \Delta \tau = \frac{(N - 1) l_{\rm arc}}{v_{\rm 0, cr}} < \tau_{\rm A}. 
\end{equation}
Here, $\tau_{\rm A} \approx 47$ s and we determined $E_{\rm cr}\approx 23$ and 17 mV/m ($v_{\rm 0, cr} \approx 0.46$ and 0.34 km/s) for $(2, 0)$ and $(3, 0)$ modes. Substituting these values into Eq.\ (6), we calculated $\Delta \tau \approx 21$ and 58 s, respectively, and the latter does not satisfy the above inequality. Thus, we are compelled to remove the possibility of the instability with higher modes $k_x \ge 3$. It is, however, noted that the $(3, 0)$ mode can be preferable rather than the $(2, 0)$ mode in the other specific case of Eq.\ (6) with a stretched magnetic field. 

For motivations of future observations of electron energy spectrum just before auroral intensification, we illustrated the way to estimate a realistic $q$ value. Equating it, 
\begin{equation}
 \displaystyle q = \eta \frac{\epsilon}{\epsilon_{\rm th}}, 
\end{equation}
which meets $q \ge 1$. Here, $\epsilon$ stands for injected electron energy (eV), $\epsilon_{\rm th}$ the mean energy loss per ion pair production, and $\eta$ the ratio of the injected (non-thermal) electron flux to the thermal electron flux carrying Alfv$\acute{\rm e}$nic $j_\parallel$. The typical value of $\epsilon_{\rm th}$ was estimated to be 35 eV [Fang et al., 2008; and references therein]. Generally, $\eta < 1$ since we assumed the ionization term in Eq.\ (3) as a product $q j_\parallel$ of $q$ and $j_\parallel$ proportional to the thermal electron flux. Referring to the precipitating electron model by Fang et al.\ [2008], we pick energies of $\epsilon = 5$ or 10 keV that produce peaks in the ionization rate at our target region of 100--120 km. For these energies, $\epsilon / \epsilon_{\rm th} \approx 140$ and $\approx 280$, respectively. We assume measured energy distributions with EISCAT radars [Kirkwood et al., 1990; Olsson et al., 1996] and DMSP satellite [Yago et al., 2005] as a 1 keV Maxwellian distribution, and roughly estimate the energy ratio $\eta$ to be up to $\approx 10^{-2}$. In this case, $q \approx 1.4$ and 2.8 are obtained for 5 and 10 keV electrons, respectively. 

The above estimation of $q$ values is supported by the assumption that an inverted-V potential right above the arc keeps a robust energy spectrum including accelerated electrons. However, it has a possibility to change the $q$ value (even a slight change) if the inertial Alfv$\acute{\rm e}$n wave supplies $< 1$ keV accelerated electrons; here, we refer to the stopping height of 1 keV electrons being $\approx 100$ km [Fang et al., 2008]. Although we did not consider any dependence of $q$ on $j_\parallel$, $q$ could correlate with $j_\parallel$ of shear Alfv$\acute{\rm e}$n waves through $E_\parallel$ of inertial Alfv$\acute{\rm e}$n waves. The linear form of $q j_\parallel$ may be corrected in our future studies that treat both waves. 

Our prime suggestion for future observations is that an increase in the ionization rate by non-thermal electrons can reduce the threshold field, to a great extent, of instabilities of Alfv$\acute{\rm e}$n waves related to auroral intensification. However, compared to the average energy component carrying $j_\parallel$, the amount of the non-thermal component is limited. The $q$ value has an upper limit, and thus $E_{\rm cr}$ value has a lower limit. SuperDARN radar observations revealed that the convection speed in the vicinity of arcs exceeds $\approx 400$ m/s over $\approx 30$ min before onsets of auroral intensification [Bristow and Jensen, 2007]; the time scale ($> \tau_{\rm A}$) is reasonable for shear Alfv$\acute{\rm e}$n waves to be amplified [see Hiraki, 2015b]. With the magnetic field of $B_0 \approx 5 \times 10^{-5}$ T, the speed corresponds to the electric field of $E_0 \approx 20$ mV/m. The rate $q = 3$--4 is sufficient for instabilities of the waves, if this field value is regarded as the threshold. However, we need the higher rate $q = 5$--6, if future precise and statistical observations will present a lower value, e.g., $\min E_{\rm cr} \approx 10$ mV/m. We urge fascinating field/particle observations to determine both the lower limit of threshold fields and the upper limit of $q$ values.

\section{Conclusion}
This study investigated changes in the threshold electric field $E_{\rm cr}$ of plasma convection, causing a linear instability of Alfv$\acute{\rm e}$n waves, due to the ionization ($q$) by precipitating auroral electrons. We clarified that the value of $E_{\rm cr}$ is reduced in the cases of $q > 1$. A realistic value of $q \approx 3$, sufficient for a fast growth of waves, was obtained from observed electron energy spectra. In future, our results motivate the thorough search for the $q$ value with electron energy spectra at auroral breakup and precise measurements of convection electric fields by radars.




%
%

\end{document}